\documentclass[aps,pra,twocolumn,superscriptaddress,amsmath,amsfonts,amssymb,preprintnumbers]{revtex4-1}
\usepackage[T1]{fontenc}
\usepackage{wasysym}
\usepackage[latin1]{inputenc}
\usepackage{graphicx}
\usepackage[colorlinks,plainpages=false,linkcolor=black,urlcolor=blue,citecolor=black,pdfpagemode=UseNone,pdfstartview=FitBH]{hyperref}

\makeatletter
\renewcommand{\@evenfoot}{\hfill\raisebox{-3em}{\bf\thepage}\hfill}
\renewcommand{\@oddfoot}{\hfill\raisebox{-3em}{\bf\thepage}\hfill}
\makeatother

\begin{document}

\title{Ballistic magnon heat conduction and possible Poiseuille flow in the helimagnetic insulator Cu$_2$OSeO$_3$}
\author{N.~Prasai}
\affiliation{Department of Physics, University of Miami, Coral Gables, FL 33124}
\author{B.~A. Trump}
\affiliation{Department of Chemistry, Johns Hopkins University, Baltimore, MD 21218}
\affiliation{Department of Physics and Astronomy, Institute for Quantum Matter, Johns Hopkins University, Baltimore, MD 21218}
\author{G.~G. Marcus}
\affiliation{Department of Physics and Astronomy, Institute for Quantum Matter, Johns Hopkins University, Baltimore, MD 21218}
\author{A.~Akopyan}
\affiliation{Department of Physics, University of Miami, Coral Gables, FL 33124}
\author{S.~X. Huang}
\affiliation{Department of Physics, University of Miami, Coral Gables, FL 33124}
\author{T.~M. McQueen}
\affiliation{Department of Chemistry, Johns Hopkins University, Baltimore, MD 21218}
\affiliation{Department of Physics and Astronomy, Institute for Quantum Matter, Johns Hopkins University, Baltimore, MD 21218}
\affiliation{Department of Material Science and Engineering, Johns Hopkins University, Baltimore, MD 21218}
\author{J.~L.~Cohn}
\email[Corresponding Author: ]{cohn@physics.miami.edu}
\affiliation{Department of Physics, University of Miami, Coral Gables, FL 33124}

\begin{abstract}
We report on the observation of magnon thermal conductivity $\kappa_m\sim 70$~W/mK near 5~K in the helimagnetic insulator
Cu$_2$OSeO$_3$, exceeding that measured in any other ferromagnet by almost two orders of magnitude. Ballistic,
boundary-limited transport for both magnons and phonons is established below 1~K, and Poiseuille flow of magnons is
proposed to explain a magnon mean-free path substantially exceeding the specimen width for the least defective specimens
in the range $2 {\rm K}<T<10$~K. These observations establish Cu$_2$OSeO$_3$ as a model system for studying long-wavelength magnon dynamics.
\end{abstract}

\maketitle
\clearpage

\section{Introduction}

Spin-mediated heat conduction in ferromagnetic materials has been of interest for decades, but a dearth of suitable ferro- or
ferrimagnetic insulators exhibiting magnonic heat conduction has limited
investigation \cite{FriedbergDouthett,McCollum,Luthi,Douglass,BhandariVerma,WaltonRives,Pan,BoonaHeremans}.
The most widely studied example is yttrium-iron garnet (YIG), for which a small magnonic thermal conductivity is well-established at
low temperatures. Magnon heat conduction and energy exchange between magnons and phonons have attracted renewed attention
recently because of their importance for the burgeoning fields of spin caloritronics \cite{SpinCaloritronics} and magnon
spintronics \cite{MagnonSpintronics} wherein thermally-driven spin currents induce electrical signals. Essential to the development
of related technologies is a deeper understanding of magnon heat conduction and magnon-phonon interactions generally, and
identifying suitable materials for realizing practical devices.

Here we report magnon thermal conductivities $\kappa_m\sim 70$~W/mK near 5~K in single crystals of the helimagnetic insulator
Cu$_2$OSeO$_3$, far exceeding those observed previously in any other ferro- or ferrimagnets (including YIG). Distinguished in applied magnetic field,
both the magnon and phonon ($\kappa_L$) thermal conductivities exhibit ballistic behavior below 1~K,
with mean-free paths (mfps) limited by the specimen boundaries and $\kappa_m\propto T^2$, $\kappa_L\propto T^3$. At $T>1$~K, $\kappa_m$
for clean specimens increases substantially faster than $\propto T^2$ and reaches values twice as large as expected from spin-wave theory. We
consider both magnon-phonon drag and Poiseuille flow of magnons as potential mechanisms for this enhancement, and
present analysis supporting the latter.

Cu$_2$OSeO$_3$ is a cubic material \cite{CrystalSymmetry,CrystalSymmetry2} (space group P2$_1$3), consisting
of a three-dimensional distorted pyrochlore (approximately fcc) lattice of corner-sharing Cu tetrahedra. Inequivalence of the
copper sites and strong magnetic interactions within tetrahedra lead to a 3-up-1-down, spin $S=1$ magnetic state \cite{Palstra,Belesi} that
persists above the long-range magnetic ordering temperature \cite{tetrahedronGS,ESR}. Weaker interactions between tetrahedra
lead to their ferromagnetic ordering below $T_C\simeq 58$~K. Dzyaloshinsky-Moriya interactions induce a long-wavelength, incommensurate helical
spin structure and promote a Skyrmion lattice phase \cite{Skyrmion,SpecHeat} near $T_C$ that has attracted considerable attention.
At low temperatures the low-field state is helimagnetic wherein the atomic spins rotate within a plane perpendicular to the helical axis
with a wavelength $\lambda_h\simeq 62$~nm; mutliple domains with helices aligned along $\langle 100\rangle$ directions characterize this phase.
At $H\gtrsim 300$~Oe the helices of individual domains rotate along the field to form a single-domain, conical phase in which spins
rotate on the surface of a cone. Further increasing the field narrows the conical angle until $H\gtrsim 1$~kOe where
the ferrimagnetic, collinear-spin state emerges.
\begin{figure*}[t]
\includegraphics[width=7in,clip]{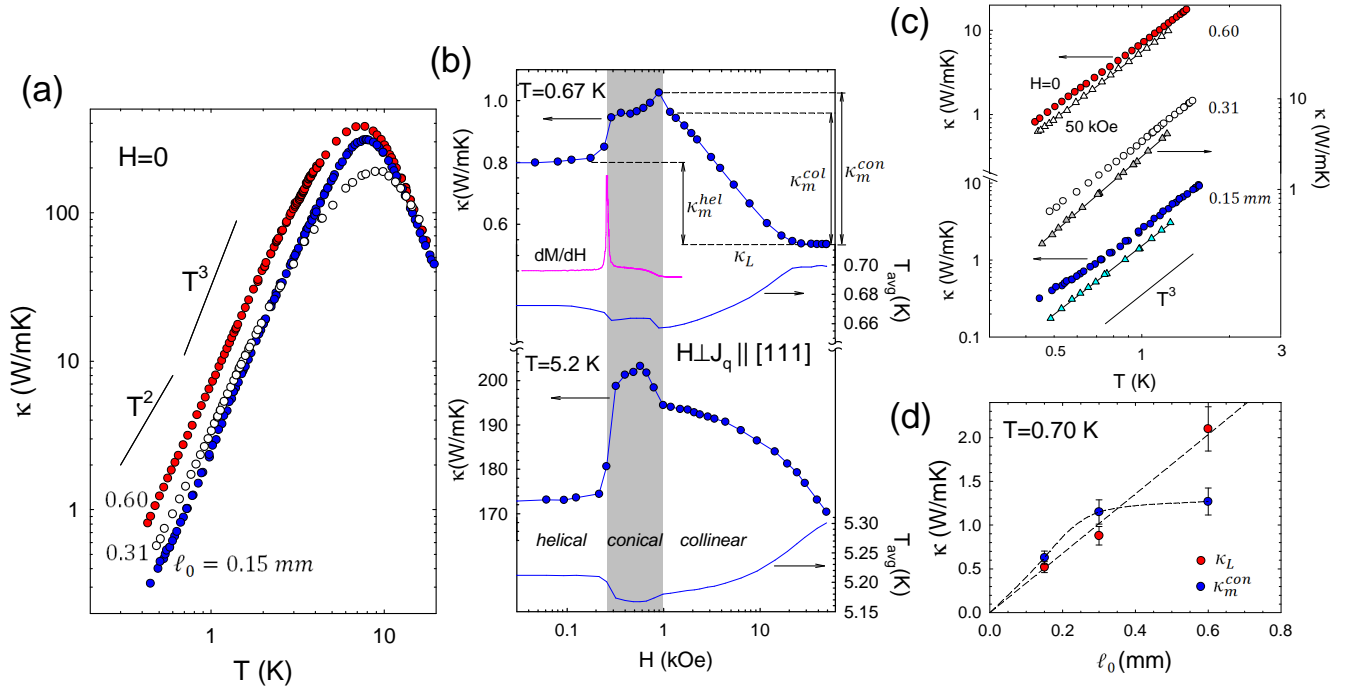}
\caption{
(a) Thermal conductivity measured along $[111]$ for three specimens (at $H=0$) labeled by their transverse dimensions $\ell_0$.
(b) Magnetic field dependence of thermal conductivity (left ordinates) and average specimen temperature (right ordinates) at two temperatures for $\ell_0=0.15$~mm. Also shown in the
upper panel is $dM/dH$ at 1.9~K. Here $H$ is the internal field, i.e. corrected for demagnetization.
The gray shading delineates the different spin phases.
The lattice contribution $\kappa_L$ is identified as the high-field saturation value of $\kappa$ for $T=0.67$~K, and $\kappa_m$ in the helical, conical, and collinear phases
as differences (vertical arrows and dashed lines)
(c) Low-T data for the same three specimens from (a) at $H=0$ (circles) and $H=50$~kOe (triangles). The solid lines are linear-least-squares fits.
(d) $\kappa_L$ and $\kappa_m^{con}$ \emph{vs.} $\ell_0$ at $T=0.70$~K for the three specimens from (a).
}
\label{Fig1}
\end{figure*}
\section{Experimental Methods}

Phase pure, single crystals of Cu$_2$OSeO$_3$ were grown by chemical vapor transport \cite{Growth}. Cu$_2$OSeO$_3$ powder was first synthesized by
three stoichiometric (2:1 CuO:SeO$_2$) heat treatments at 600 $^{\circ}$C, each followed by quenching and grinding. The resulting powder was
placed in an evacuated fused-silica tube with a temperature gradient of 640 $^{\circ}$C - 530 $^{\circ}$C, with NH$_4$Cl as the transport additive.
After six weeks, single crystals with typical sizes of 75-125 mm$^3$ were seen, and seed crystals were also added to increase yield. Purity of single
crystals were verified by magnetization and X-ray diffraction experiments, showing reproducibility of physical property behavior and good crystallinity.

Specimens were cut from single-crystal ingots, oriented by x-ray diffraction, and polished into thin parallelopipeds. We focus in this work on specimens
with heat flow along the $[111]$ direction and perpendicular magnetic field applied along $[1\bar{1}0]$ for which our data is most extensive. Data for other
orientations of heat flow and applied field will be presented elsewhere \cite{PrasaiUnpub}. A two-thermometer, one-heater method was employed to measure the thermal conductivity in applied magnetic
fields up to 50 kOe. Specimens were suspended from a Cu heat sink with silver epoxy and affixed with a 1 k$\Omega$ chip heater on the free end.
A matched pair of RuO bare-chip sensors, calibrated in separate experiments and mounted on thin Cu plates, were attached
to the specimen through 0.125~mm diameter Au-wire thermal links bonded to the Cu plates and specimen with silver epoxy.
Measurements were performed in a $^3$He ``dipper'' probe with integrated superconducting solenoid.

A total of 5 different crystals were studied with transverse dimensions, $\ell_0\equiv 2\sqrt{{\rm\emph{\large a}}/\pi}$ (${\rm\emph{\large a}}$ is the cross-sectional area)
ranging from 0.15-0.60~mm. Three of these ($\ell_0=0.15, 0.31, 0.60$~mm) are the primary focus of this work. A fourth crystal for which data is less complete, was
cut from the same ingot as $\ell_0=0.15$~mm and appears in Fig.~\ref{Fig2}. Data for the fifth crystal appears in Appendix D, Fig.~\ref{Fig7}.

\section{Results and Discussion}
\subsection{Zero-field thermal conductivity}

Figure \ref{Fig1}~(a) shows $\kappa(T)$ for $H=0$ on three
crystals labeled by their transverse dimension ($\ell_0$). Notable
is the magnitude which reaches $\sim 400$~W/mK (for
$\ell_0=0.60$~mm) at the maximum near $T=8$~K, exceptional for a
complex oxide. $\kappa$ is also strongly sample dependent for
$T<10$~K, scaling with $\ell_0$ at the lowest $T$, but not in the
region of the maxima. As we discuss further below, the latter
feature is attributable to differing point-defect concentrations
to which $\kappa_L$ is sensitive near its maximum. Here we note
the likely defects are Se vacancies (common in Se compounds
\cite{SeDefects}) and numerical modeling of $\kappa_L$ (Appendices
D, E, Fig.~\ref{Fig6}) implies vacancy concentrations per f.u. of
5.6$\times 10^{-4}$, $1.6\times 10^{-3}$, and $4.1\times 10^{-3}$
for the specimens with $\ell_0=0.15$~mm, 0.60~mm, and 0.31~mm,
respectively.

We assume the measured thermal conductivity to be a sum of lattice (phonon) and magnon
contributions, $\kappa=\kappa_L+\kappa_m$, valid in the boundary scattering regime ($T\lesssim 3$~K as discussed below) when the phonon-magnon
relaxation time ($\tau_{ph-m}$) exceeds, but is comparable to, the phonon-boundary scattering time ($\tau_b$) \cite{SandersWalton}. Assuming the $q=0$
relaxation to be representative of the magnon system, an estimate, $\tau_{ph-m}\sim 3\times 10^{-8}$~s at 30~K, can be inferred from intrinsic ferromagnetic
resonance linewidths \cite{Seki}. Since the magnon density declines as $T^{3/2}$, $\tau_{ph-m}$ should increase to $\sim 10^{-7}-10^{-6}$~s at $T\lesssim 3$~K
where $\tau_b=\ell_0/v_{ph}\sim 10^{-7}$~s (using $v_{ph}\approx 2$~km/s); thus the assumption is justified.

\subsection{Ballistic lattice and magnon thermal conductivities distinguished in applied field}

The magnetic field dependence of $\kappa$ through the various spin phases [Fig.~\ref{Fig1}~(b)], allows
for distinguishing $\kappa_L$ and $\kappa_m$. The key features of $\kappa(H)$: (1) abrupt changes of $\kappa$ at the phase boundaries,
(2) a suppression of $\kappa$ with increasing field in the collinear phase and saturation at the highest fields (50~kOe) and lowest $T$. Behavior (2) is typical of $\kappa_m$
in ferro- and ferrimagnets \cite{McCollum,Luthi,Douglass,BhandariVerma,WaltonRives,Pan,BoonaHeremans} -- spin-wave excitations are depopulated (gapped)
for fields such that ${\textsl g}\mu_B H\gg k_BT$ (Fig.~\ref{Fig4} in Appendix A shows that the field at which $\kappa(H)$ saturates corresponds to
${\textsl g}\mu_B H/k_BT\simeq 6$). With \cite{g-factor} ${\textsl g}\simeq 2.1$ the magnon gap is $\sim 0.14$~K/kOe, such that
$\kappa(50\ {\rm kOe})\simeq \kappa_L$ for $T\lesssim 1.2$~K.

We find $\kappa_L\propto T^n$ [triangles, Fig.~\ref{Fig1}~(c)] with $n=2.7-3$,
consistent with phonon mfps limited by the specimen boundaries [Fig.~\ref{Fig1}~(d)] and nearly diffuse scattering.
\begin{figure}[t]
\includegraphics[width=3.4in,clip]{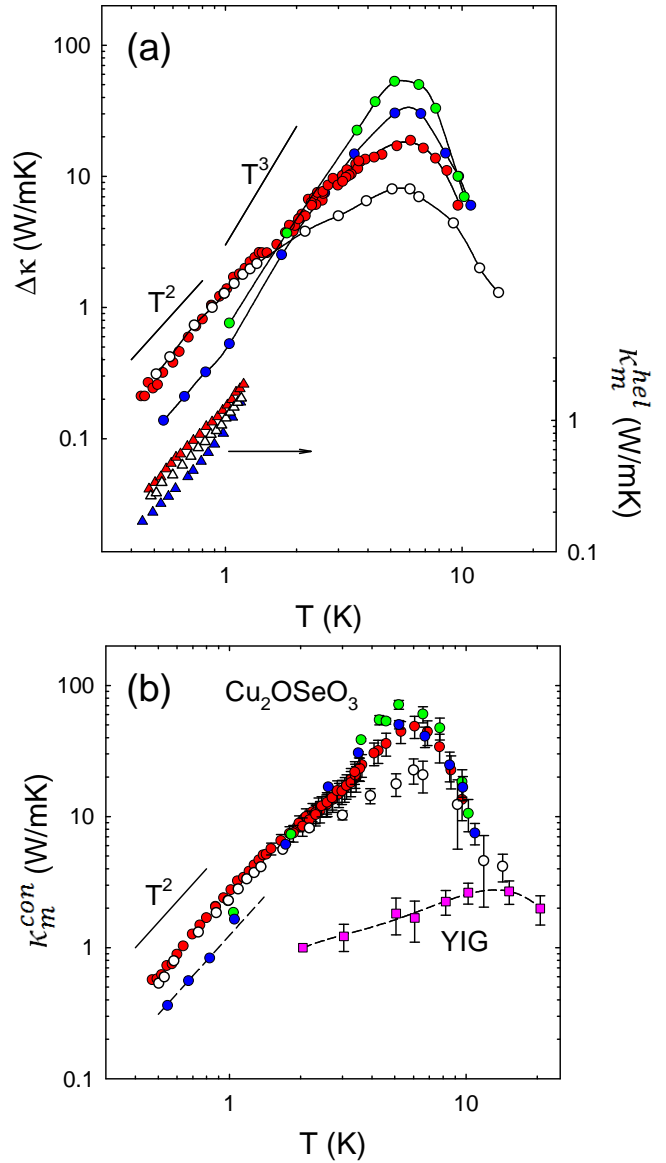}
\caption{ (a) $\kappa_m^{hel}=\kappa(H=0)-\kappa(H=50 {\rm kOe})$
(triangles, right ordinate) and
$\Delta\kappa=\kappa_m^{con}-\kappa_m^{hel}$ (circles, left
ordinate). (b) $\kappa_m$ in the conical phase (circles) for the
specimens from (a). Error bars reflect uncertainties in the
determination of $\kappa_L$ from the Callaway model (Appendix D,
Fig.~\ref{Fig6}). Also shown are $\kappa_m$ data for YIG (squares)
from Ref.~\onlinecite{BoonaHeremans}.} \label{Fig2}
\end{figure}
The Casimir expression for diffuse scattering, boundary-limited thermal conductivity can be used to determine the phonon mean-free path ($\ell_{ph}$) \cite{Casimir},
\begin{equation}
\nonumber 
  \kappa_L=\left(\frac{2\pi^2}{15}\right)\left(\frac{k_BT}{\hbar}\right)^3k_B\langle v^{-2}\rangle \ell_{ph}
\end{equation}
\noindent
where $\langle v^{-2}\rangle=[(1/3)(1/v_{LA}^3+2/v_{TA}^3)]^{2/3}$ is the Debye averaged sound velocity.
A fit of the low-$T$ $\kappa_L(T)$ data [Fig.~1~(c)] to the form $\kappa_L=AT^n$ yields $A=1.52, 2.32, 5.62$ and $n=2.96, 2.80, 2.70$,
respectively, for the specimens with $\ell_0=0.15, 0.31, 0.60$~mm. The power of $T$ slightly less than 3 is common in insulators \cite{Ziman},
indicating some specularity to the boundary scattering.
Consistent with observations, the $\ell_0=0.60$~mm specimen ($n=2.70$) was polished on one of its large faces with finer
abrasive (1~$\mu{\rm m}$) than the other specimens (5~$\mu{\rm m}$).
Longitudinal and traverse sound velocities for the $[111]$ direction from ultrasonic measurements \cite{Inosov} are $v_{LA}\simeq 3.3$~km/s
and $v_{TA}\simeq 1.85$~km/s, respectively.
Combining these parameters in the above equation yields $\ell_{ph}\simeq 0.16, 0.24, 0.59$~mm, in good agreement with the effective transverse
dimension of the specimens.

The corresponding $\kappa_m$ in the helical and conical phases computed by subtraction [vertical arrows and dashed lines, Fig.~\ref{Fig1}~(b)],
are $\propto T^2$ for $T\leq 1$~K, consistent with constant magnon mfps (Fig.~\ref{Fig2}; $\kappa_m^{col}$ is omitted for clarity).
For boundary-limited spin-wave heat conduction we have \cite{FriedbergDouthett},
\begin{equation}
\nonumber 
  \kappa_m=\frac{\zeta(3)k_B^3\ell_m}{4\pi^2\hbar D}T^2,
  \end{equation}
\noindent where $\zeta(3)\simeq 1.202$. A fit of the $\kappa_m^{con}(T)$ data [Fig.~2~(b)] at $T< 1$~K to the
form $BT^2$ gives $B=1.25, 2.3, 2.6$~W/mK$^3$, respectively, for the specimens with $\ell_0=0.15, 0.31, 0.60$~mm; the equation above implies
$\ell_m\simeq 0.14, 0.25, 0.28$~mm.  The value of $\ell_m$ for the $\ell_0=0.60$~mm specimen is significantly smaller than the specimen dimension,
suggesting a maximum magnetic domain size. Similarly, a value of $\ell_m\sim 0.34$~mm for this specimen is inferred from a plot of
$\kappa_m$ {\it vs} $\ell_0$ [Fig.~1~(d)]. Within the multi-domain helical phase, values for $\ell_m$ are roughly half as large.

The ballistic character of the magnon transport in the $T^2$
regime is further corroborated by using kinetic theory to convert
$\kappa^{con}_m$ (or $\kappa^{col}_m$) to magnetic specific heat
($C_m$) and then comparing the latter to expectations of spin-wave
theory. We have $C_m=3\kappa_m/(v_m\ell_m)$, where
$v_m=(2/\hbar)Dq$, $D=52.6$~meV\ \AA$^2$ is the spin-wave
stiffness \cite{Portnichenko} (the dispersion at low energy is
well-described \cite{DipoleNote} by $E=Dq^2$). The dominant
magnons for boundary-limited $\kappa_m$ have
\cite{dominantmagnons} $q_{dom}=(2.58k_BT/D)^{1/2}$ such that
$v_m\simeq 1040 T^{1/2}$~m/s. Assuming diffuse scattering of
magnons at the crystal (or domain) boundaries, the computed $C_m$
for all crystals agrees well with linear spin-wave theory
(Appendix~B, Fig.~\ref{Fig5}).

A transfer of energy from the spin system to the lattice as the magnon gap opens is implied, given the near-adiabatic conditions
of the specimens during measurement. The corresponding increase in the average temperature of the sample ($T_{avg}$) in the high-field regime
[solid curves, right ordinates in Fig.~\ref{Fig1}~(b)] should reflect only a fraction of the total spin energy, since much of it must be
distributed within thermometers, thermal links, and heater. As a further self-consistency check
on our analysis, this fraction is determined (Appendix~C) to be $\sim 4\%\ (30\%)$ at $T=0.67$~K ($5.2$~K).

\subsection{Determining the magnon thermal conductivites at higher $T$}

Given that the phonon mfps are boundary-limited at $T\leq 1$~K, the abrupt \emph{increase} in $\kappa$ at the helical-conical
transition [$H\approx 250$~Oe in Fig.~\ref{Fig1}~(b)] must be attributed to an increase in $\kappa_m$ associated with the approximate
doubling of $\ell_m$ noted above. It is significant that this jump, $\Delta\kappa=\kappa_m^{con}-\kappa_m^{hel}$ [Fig.~\ref{Fig2}~(a)],
exhibits the same $\propto T^2$ behavior for magnon boundary scattering at low $T$ as found for both $\kappa_m^{hel}$ and $\kappa_m^{con}$
computed by subtracting $\kappa_L$ (Fig.~\ref{Fig2}). Since $\Delta\kappa$ is independent of any assumptions regarding $\kappa_L$, it
validates the implicit assumption that $\kappa_L$ is independent of field.

At $T>1.2$~K where the applied field is insufficient to fully suppress $\kappa_m$, $\Delta\kappa$
represents a \emph{lower bound} on $\kappa_m^{con}$ [Fig.~\ref{Fig2}~(b)] since we expect $\kappa_L<\kappa_m^{hel}$
as is clear in the data of Fig.~\ref{Fig1}~(b) at $T=5.2$~K.  Very similar results for $\Delta\kappa(T)$ were found for a
specimen with $[110]$ heat flow and perpendicular field along $[1\bar{1}0]$, thus a large $\kappa_m$ is not restricted
to the $[111]$ direction \cite{PrasaiUnpub}. The sharp decline of $\Delta\kappa$ at $T\gtrsim 7$~K, and its disappearance
for $T\gtrsim 12$~K, indicate that $\kappa_m$ has a maximum at $T\sim 5-6$~K and becomes negligible for $T\gtrsim 12$~K.
The latter is supported by recent spin-Seebeck measurements \cite{Aqeel} indicating a sharp decline in spin-polarized heat
current in the same temperature regime.

To estimate $\kappa_m^{con}$ at higher $T$, this behavior of
$\kappa_m$ and the low-$T$ $\kappa_L$ are exploited as strong constraints on calculations of $\kappa_L(T)$ at $T\geq 1.2$~K
using the Callaway model (Appendix~D, Fig.~\ref{Fig6}). This procedure, dictates the error bars on $\kappa_m^{con}$
in Fig.~\ref{Fig2}~(b) and, as noted above, provides estimates of specimen defect (Se vacancy) concentrations (Appendix~E).
\begin{figure*}[t]
\includegraphics[width=6.in,clip]{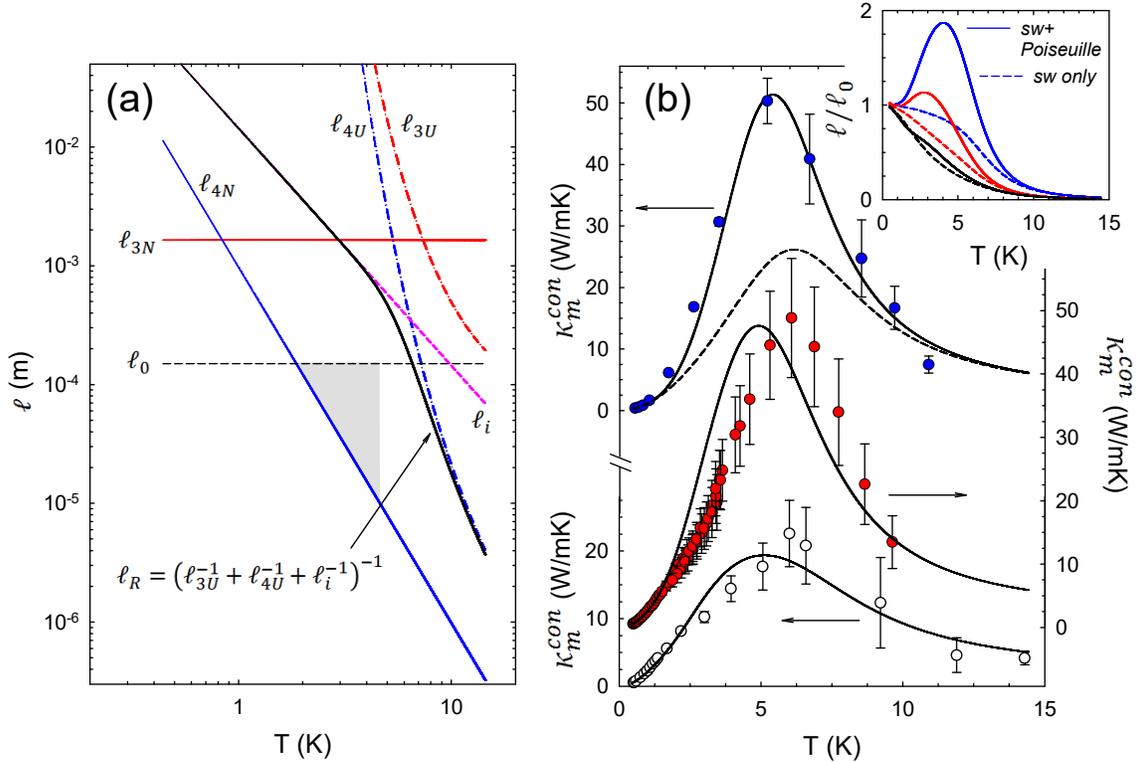}
\caption{ (a) Magnon mean-free paths for scattering from the model
of Ref.~\onlinecite{ForneyJackle} (see Appendix F for details).
Subscripts refer to 3-magnon and 4-magnon normal (3N, 4N) and
umklapp (3U, 4U) processes, elastic impurity scattering (i), and
total resistive scattering (R). The Poiseuille conditions (see
text) are met in the shaded region. (b) $\kappa_m^{con}(T)$ for
the three crystals from Fig.'s 1 and 2 with linear scaling. The
solid curves are model predictions for elastic defect
concentrations (from top to bottom in ppm): 12, 22, 62. The
dashed curve for the $\ell_0=0.15$~mm specimen represents the
spin-wave contribution alone without Poiseuille enhancement.
Inset: magnon mfps from the model, normalized by low-$T$
boundary-limited values, for each specimen.} \label{Fig3}
\end{figure*}

\subsection{Anomalous $T$ dependence for $\kappa_m$ and possible Poiseuille flow}

A most striking feature of both $\Delta\kappa(T)$ and $\kappa_m^{con}(T)$, aside from unprecedented magnitudes, is their increase, for the two least defective specimens,
with a substantially higher power of $T$ than $\propto T^2$ at $T\geq 1$~K (Fig.~\ref{Fig2}).
An additional contribution to $C_m$ from spin-wave "optic" modes cannot be expected in this temperature regime since those
sufficiently dispersive to contribute to $\kappa_m$ have energies exceeding $25$~meV \cite{Portnichenko}.
We are aware of only two possible mechanisms that can potentially explain this observation: (1) magnon-phonon drag, (2) Poiseuille flow of magnons.
Theory suggests that for momentum-independent magnon relaxation time $\tau_m$, an additive phonon-magnon drag contribution should take the general form \cite{Chernyshev},
$\kappa_{drag}\sim (1/3)C_Lv_m^2\tau_m\propto T^4\tau_m$, thus offering a stronger $T$ dependence. The relevant magnon-phonon interactions are normal,
momentum-conserving processes.

A more intriguing alternative is that magnons undergo Poiseuille flow, predicted 50 years ago for both phonons and
magnons \cite{Gurzhi,GuyerKrumhansl,ForneyJackle}, but observed only for phonons and only in exceptionally clean materials
(e.g. crystalline $^4$He \cite{He4}). When the mfp for normal scattering ($\ell_N$) is much shorter
than both the transverse dimension ($\ell_0$) and the mfp for bulk resistive scattering processes ($\ell_R$), quasiparticles
undergo many momentum-conserving scattering events before losing their momentum at the specimen boundaries. Under the stringent
conditions $\ell_N<\ell_0/2< (\ell_N\ell_R)^{1/2}$, the effective mfp approaches that
for a particle undergoing random walk with step size $\ell_N$, mfp$\sim \ell_0^2/4\ell_N\gg\ell_0$. We pursue this scenario further
since all of the relevant scattering rates for magnons have been computed \cite{Akheizer,ForneyJackle} for a Heisenberg ferromagnet in
the low-$T$ regime, and interactions
with phonons which underlie phonon-drag are predicted to be significantly weaker.

Forney and J\"ackle \cite{ForneyJackle} calculated rates for normal and umklapp magnon scattering and elastic magnon-impurity scattering
(non-magnetic defects). The
expressions contain three parameters (Appendix~F), two of which are set by the lattice constant and exchange coupling. The only remaining
free parameter is the defect concentration.
Figure~\ref{Fig3}~(a) shows the relevant mfp's employed for the least defective crystal ($\ell_0=0.15$~mm). The conditions for Poiseuille flow
are met in the shaded region.
$\kappa_m^{con}$ is computed [solid curves, Fig.~3~(b)] from the kinetic theory expression with a mfp described by an
interpolation formula [eq.~(F1)] that yields
the conventional resistive scattering
length well outside the Poiseuille window, $\ell_R^B=(1/\ell_0+1/\ell_{3U}+1/\ell_{4U}+1/\ell_i)^{-1}$, and tends toward $\ell_0^2/\ell_N$
within the Poiseuille regime. Interpolation is controlled by ``switching factors'' \cite{GuyerKrumhansl,deTomas} related to the ratio
$\ell_N/\ell_R$ (Appendix F and Fig.~\ref{Fig8}). The data are well-described by the model
(with defect concentrations 12, 22, 62 ppm for $\ell_0=0.15$~mm, 0.60~mm, and 0.31~mm), though
the computed maxima for the more defective specimens deviate from experiment, a consequence of the Poiseuille window being shifted to
lower $T$ as the impurity scattering
mfp decreases. This may signal inadequacy of the magnon-impurity scattering model, perhaps because spin defects in the present system
may be associated with Se vacancies
as suggested by a correlation between the defect concentrations inferred for magnons and phonons (Appendix F, Fig.\ref{Fig9}).

\section{Summary}

Our observations reveal Cu$_2$OSeO$_3$ to be a model system for further study of long-wavelength magnon dynamics, e.g. our proposal that
magnons undergo Poiseuille flow implies that magnon ``second sound'' might also be observed.
Since both the conical and collinear-phase magnon heat conductivities are similar in magnitude, helical magnetism is evidently not the origin
of its unusually large $\kappa_m$. Since long-wavelength magnons play a prominent role in the
spin-Seebeck effect \cite{BoonaHeremans,SpinCaloritronics} the results presented
here also make it possible to investigate interfacial spin-current transfer using calibrated magnon heat currents and to explore
the possible role of the spin phases on transfer efficiency.

\section{Acknowledgments}

The authors acknowledge helpful comments from A.~L.~Chernyshev. This material is based upon work supported by U.S.
Department of Energy (DOE), Office of Basic Energy Sciences (BES) Grant No.'s DEFG02-12ER46888 (University of Miami) and
DEFG02-08ER46544 (Johns Hopkins University).

\begin{figure*}[t]
\includegraphics[width=5.5in,clip]{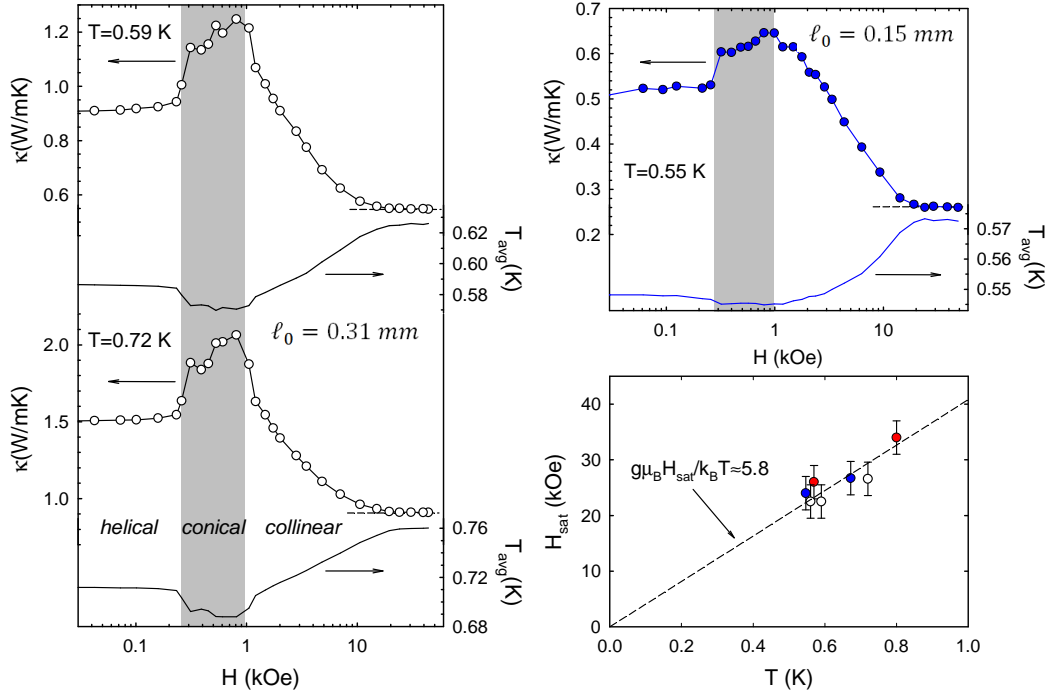}
\caption{Magnetic field dependence of thermal conductivity (left
ordinates) and average specimen temperature (right ordinates) for
$\ell_0=0.31$~mm (left panel) and $\ell_0=0.15$~mm (upper right
panel). The lower right plot shows the field at which $\kappa$ becomes
field-independent, $H_{sat}$ {\it vs.} $T$ for all three crystals
at the lowest $T$. Symbols employed are the same as those from Fig.'s~1-3.}
\label{Fig4}
\end{figure*}

\appendix

\section{Additional low-$T$ $\kappa(H)$ data}

Figure~\ref{Fig4} shows additional low-$T$ $\kappa(H)$ data showing suppression of the magnon contribution at high fields where we infer $\kappa=\kappa_L$. We also
plot the field $H_{sat}$ at which $\kappa$ becomes field-independent against temperature.

\section{Magnetic specific heat computed from $\kappa_m$}

As noted in Ref.~\onlinecite{Portnichenko}, the Cu$_4$ tetrahedra of Cu$_2$OSeO$_3$ approximate an fcc lattice, the primitive cell
of which is 4 times smaller than that of the simple cubic cell. Thus the standard low-temperature form of the magnetic specific
heat per volume becomes, $C_m=(0.113/4) k_B\left(k_B T/D\right)^{3/2}$ (this factor of $1/4$ also appears in
expressions for the spin-wave thermal conductivity). Values of $C_m$ (as described in the text) were computed from the
measured $\kappa_m^{col}$ (or $\kappa_m^{con}$) using kinetic theory and $\ell_m=\ell_0$ for the four crystals from Fig.~2~(a),
with the exception of the $\ell_0=0.60$~mm crystal for which we used $\ell_m=0.34$~mm based on the effective length inferred from Fig.~1~(d).
Theory and experiment agree well (Fig.~\ref{Fig5}).

\section{Energy transfer from spins to lattice at high field}

We estimate the fraction of total spin energy transferred to the lattice of the $\ell_0=0.15$~mm specimen at $T=0.67$~K, upon gapping out the spin
waves in maximum field [Fig.~1~(b)], as $Q/u_m$ where $Q=C_L\Delta T$ is the heat transferred per volume, computed from the lattice
specific heat ($C_L$) and change in $T_{avg}$ induced by applied field ($\Delta T$), and $u_m$ is the total energy per volume in the spin system,

\begin{figure}[b]
\includegraphics[width=2.9in,clip]{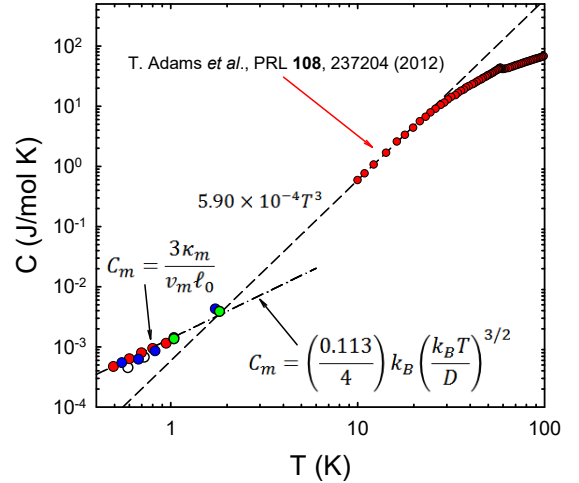}
\caption{Measured total specific heat from Ref.~\onlinecite{SpecHeat} and computed magnetic
specific heat from kinetic theory using $\kappa_m$ and effective transverse
sample dimension $\ell_0$ (specimen symbols are the same as those from Fig.'s~1-3).
The dashed line is a $T^3$ fit to the measured specific heat data at $T< 20$~K, and the dash-dotted line represents
the specific heat for an fcc magnetic
lattice, converted to molar units for Cu$_2$OSeO$_3$ using 1~mol=$5.35\times 10^{-5} {\rm m}^3$.}
\label{Fig5}
\end{figure}
\begin{equation}
\nonumber 
  u_m=\frac{D}{16\pi^2}\left(\frac{k_BT}{D}\right)^{5/2}\Gamma(5/2)\zeta(5/2;1),
\end{equation}
\noindent $\Gamma(5/2)=3\pi^{1/2}/4$ and $\zeta(5/2;1)\simeq
1.341$. With $\Delta T=0.043$~K (Fig.~1b) and using the $T^3$ fit
to the measured specific heat (dashed line, Fig.~\ref{Fig5}) to compute
$C_L$, we find $Q=0.14$~J/${\rm m}^3$ and $u_m\simeq 3.8$~J/${\rm
m}^3$, such that $Q/u_m\approx 0.036$. At $T=5.2$~K a similar analysis yields $Q/u_m\approx 0.30$.

\begin{figure}[b]
\includegraphics[width=2.9in,clip]{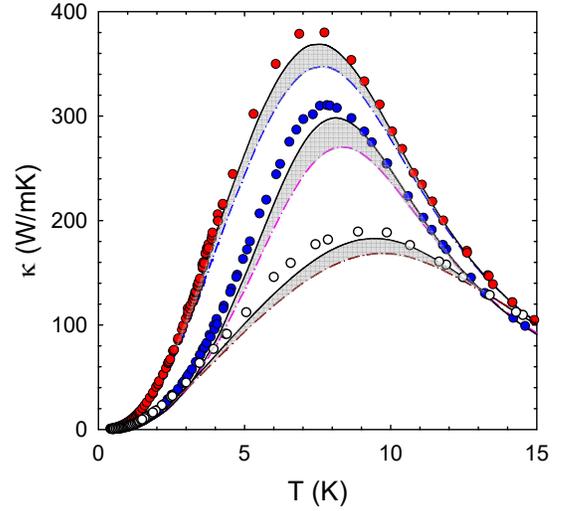}
\caption{$\kappa(H=0, T)$ for the three specimens shown in Fig.~1 (solid circles) and computed $\kappa_L$ (solid and dash-dotted curves) for two parameter sets for each specimen.
Solid curves (from top to bottom, with the same units of Table~I): $v=2.06$, $A=1.87$, $b=6.35$, $C=36$, $\gamma=1/100$; $v=2.06$, $A=1.9$, $b=6.76$, $C=10$, $\gamma=1/50$;
$v=2.35$, $A=1.72$,$b=6.35$, $C=90$, $\gamma=0$. Dash-dotted curves (from top to bottom): $v=2.15$, $A=1.77$, $b=6.6$, $C=37$, $\gamma=1/100$; $v=2.06$, $A=1.87$, $b=6.7$,
$C=14.5$, $\gamma=0$; $v=2.35$, $A=1.5$,  $b=6.35$, $C=110$, $\gamma=0$.}
\label{Fig6}
\end{figure}
\section{Calculations of $\kappa_L(T)$}

The Callaway model \cite{Berman}, incorporating its recent update
\cite{Allen}, was employed to compute $\kappa_L(T)$ for each of
the crystals, with parameter ranges restricted by the following
constraints: (1) $\kappa_L$ fits the low-$T$, high-field data
(where $\kappa_L$ is inferred directly) and the $T\geq 12$~K,
zero-field data (where $\kappa_m$ is inferred to be negligible by
the vanishing of $\Delta\kappa$), (2) the maximum in
$\kappa_{con}$, computed by subtracting $\kappa_L$ from $\kappa$
measured at the conical-collinear transition, should occur at
$T\approx 5-6$~K where $\Delta\kappa$ has its maximum, (3)
$\kappa_L\lesssim\kappa(H=50 {\rm kOe})$.

The integral expression for $\kappa_L$ is,
\begin{figure*}[t]
\includegraphics[width=5in,clip]{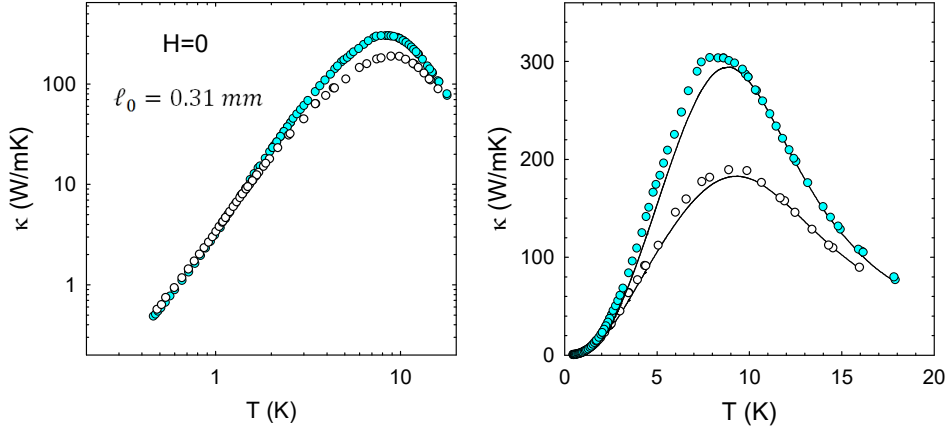}
\caption{$\kappa(H=0, T)$ for two crystals with $\ell_0\simeq 0.31$~mm in log-log scaling (left) and linear scaling (right). Open circles are for the same crystal from Fig.'s~1-3.
Solid curves are produced using the Callaway model using (in units from Table~I) $v=2.35$, $A=1.72$, $\gamma=1/50$ and: (upper curve)  $b=6.4$, $C=26$; (lower curve) $b=6.2$, $C=80$.}
\label{Fig7}
\end{figure*}
\begin{widetext}
\begin{equation*}
\nonumber 
  \kappa_{L}=\frac{k_B}{2\pi^2v}\left(\frac{k_B}{\hbar}\right)^3T^3\left[\int_0^{\Theta_D/T} \frac{x^4e^x}{\left( e^x-1\right)^2}\tau_C(x,T)dx\right]\left(1+\frac{\overline{\tau_C(x,T)/\tau_N(x,T)}}{\overline{\tau_C(x,T)/\tau_R(x,T)}}\right),
\end{equation*}
\begin{equation*}
\nonumber 
{\rm with}\quad \overline{f(T)}=\int_0^{\Theta_D/T} \frac{x^4e^x}{\left( e^x-1\right)^2}f(x,T)dx \Bigg/ \int_0^{\Theta_D/T} \frac{x^4e^x}{\left( e^x-1\right)^2}dx,
\end{equation*}
\end{widetext}

\noindent
where $v$ is the Debye averaged sound velocity (see above), $\Theta_D=(\hbar v/k_B)(6\pi^2N/V)^{1/3}$ the Debye temperature, $x=\hbar\omega/k_BT$ the reduced phonon energy,
$\tau^{-1}_C(x,T)=\tau^{-1}_N(x,T)+\tau^{-1}_R(x,T)$, and $\tau^{-1}_N(x,T)$ and $\tau^{-1}_R(x,T)$ are phonon scattering rates for normal (momentum conserving) and
resistive (momentum non-conserving) processes, respectively. $\tau^{-1}_R(x,T)$ included terms for scattering from boundaries, other phonons (Umklapp scattering), and
point-like defects (Rayleigh),
\begin{equation}
\nonumber 
  \tau_R^{-1}(x,T)=v/\ell_{ph}+Ax^2T^4\exp\left(-\frac{\Theta_D}{bT}\right)+Cx^4T^4,
\end{equation}
\noindent
where $\ell_{ph}=\ell_0$ is the boundary-limited phonon mean-free path and $A$, $b$, $C$ are constants. The normal scattering rate was
taken to have the same frequency dependence as for Umklapp scattering \cite{Allen}, but without the exponential $T$ dependency,
$\tau_N^{-1}(x,T)=\gamma Ax^2T^4$, with $\gamma$ a constant. A broad range for $\gamma$ was explored in the fitting and it was found
that only for $\gamma\leq 1/50$ were the constraints satisfied. $\gamma=1/50$ implies a normal scattering rate that begins to exceed
that for Umklapp scattering at $T\lesssim 10$~K. Phonon-magnon scattering was assumed to be substantially weaker than other scattering.

Fig.~\ref{Fig6} shows $\kappa(H=0,T)$ data for the three specimens
from Fig.~1 along with two $\kappa_L$ curves for each (solid and
dash-dotted curves). These curves border the ranges (shading)
defined by the constraints noted above. Data points for $\kappa_m$
in Fig.~2~(b) correspond to the middle of these ranges with error
bars equal to the width of the shaded region. A summary of the
scattering parameters is provided in Table~I.

In Fig.~\ref{Fig7} we compare $\kappa(T)$ at $H=0$ for the most defective $\ell_0=0.31$~mm specimen from Fig.'s~1-3 with a less
defective crystal having the same $\ell_0$. Consistent with expectations, Callaway-model parameter sets for
$\kappa_L$ (solid curves, right panel) differ principally in the defect concentration ($C$).

\vspace{-.22in}
\section{Estimate of Se vacancy concentration from point-defect fitting parameters for $\kappa_L$}

Interpreting the point-defect phonon scattering rate (Table~I above) as entirely due to Se vacancies, the vacancy concentration
can be estimated using \cite{RatsKlem},
\begin{equation}
\nonumber 
\tau_d^{-1}=\frac{n}{7}\frac{9a^3}{4\pi
v^3}\left(\frac{M_{Se}}{\overline{M}}\right)^2\omega^4,
\end{equation}
\noindent where $n$ is the concentration of vacancies on the Se
sublattice, $a=1.22$~{\AA} is the Se atomic radius, $v=2060$~m/s
is the sound velocity, and $M_{Se}/\overline{M}\simeq 2.05$ is the
ratio of the Se mass to the average mass. Using values $C=13, 37,
95\ {\rm K}^{-4}$ from Table~I for the $\ell_0=0.15, 0.60,
0.31$~mm crystals yields concentrations per f.u., $5.6\times
10^{-4}$, $1.6\times 10^{-3}$, $4.1\times 10^{-3}$, respectively.

\begin{table}[h]
\caption{Ranges of scattering parameters from Callaway modeling of $\kappa_L$.}
\begin{ruledtabular}
\begin{tabular}{cccccccc}
$\ell_0$ (mm)  & $v (\rm km/s)$ & $A (10^4\ {\rm K}^{-4})$ & $b$ & $C ({\rm K}^{-4})$ \\
0.15& $2.06-2.15$ & $1.8-2.0$ & $6.6-6.9$ & $10-15$ \\
0.31& $2.15-2.35$ & $1.5-1.8$ & $6.0-6.6$ &  $80-110$  \\
0.60& $2.06-2.3$ & $1.75-2.0$ & $6.3-6.6$ &  $34-40$  \\
\end{tabular}
\end{ruledtabular}
\end{table}

\section{Magnon scattering rates and modeling of Poiseuille flow}

Forney and J\"ackle \cite{ForneyJackle} computed the thermally
averaged 3-magnon and 4-magnon normal ($3N$, $4N$) and umklapp
($3U$, $4U$) scattering rates and magnon-impurity scattering rate
($i$) for a quadratic magnon dispersion within the Born
approximation, valid for small impurity concentration, $T\ll T_C$
and $T\gg k_B\Delta$, where $\Delta$ is the energy gap ($\sim 12\
\mu$eV for Cu$_2$OSeO$_3$):
\begin{widetext}
\begin{equation}
\nonumber 
  \tau_{3N}^{-1}=2.6S\frac{k_B}{\hbar}T_d^2 T_e^{-3/2}T^{1/2},\qquad  \tau_{4N}^{-1}=6.1\times 10^{-4} \frac{k_B T^4}{S^2\hbar T_e^3},
\end{equation}
\begin{equation}
\nonumber 
   \tau_{3U}^{-1}=1.4\times 10^3 \frac{Sk_B T_d^2}{\hbar (T_e T)^{1/2}} \exp{(-12T_e/T)},\qquad   \tau_{4U}^{-1}=\frac{2}{S^2} \frac{k_B T^{3/2}}{\hbar T_e^{1/2}}\exp{(-12T_e/T)},\qquad
 \tau_i^{-1}=0.4c\frac{k_B}{\hbar} \frac{k_B T^{5/2}}{T_e^{3/2}},
\end{equation}
\indent where
\begin{equation}
\nonumber 
  T_d=\frac{({\textsl g}\mu_B )^2}{k_B a^3},\qquad T_e=\frac{2SJ}{k_B}.
\end{equation}
\end{widetext}

\begin{figure}[t]
\includegraphics[width=2.8in,clip]{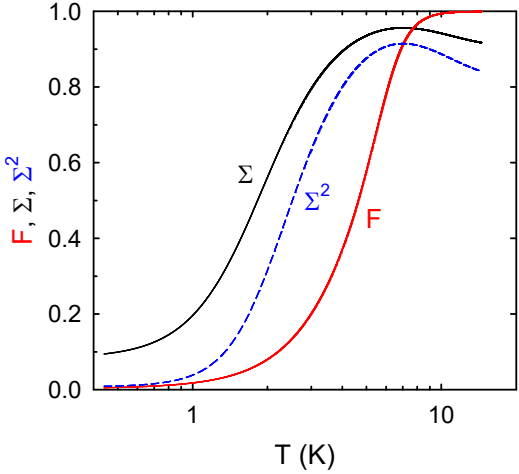}
\caption{$\Sigma$, $\Sigma^2$, and $F(L_{eff})$ used for the Poiseuille analysis of $\kappa_m^{con}$ for the least defective ($\ell_0=0.15$~mm) crystal [Fig.~3~(b)].}
\label{Fig8}
\end{figure}
\begin{figure}[t]
\includegraphics[width=2.8in,clip]{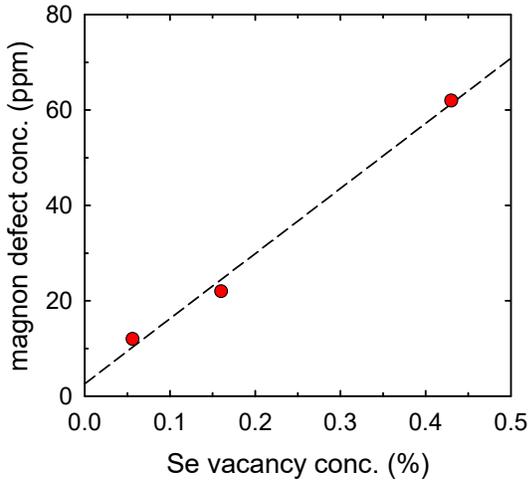}
\caption{Nonmagnetic defect concentration for magnons from the model fitting {\it vs.} Se vacancy concentration inferred from Callaway fitting of $\kappa_L$.}
\label{Fig9}
\end{figure}

We initially re-scaled the values $T_d=0.012$~K and $T_e=1.0$~K employed in Ref.~\onlinecite{ForneyJackle} for EuS ($T_C=16.5$~K) using the ratio of lattice
constants and $T_C$ (as a surrogate for $J$). These gave $T_d=0.004$~K and $T_e=3.5$~K. Subsequently we settled on $T_e=4.2$~K which
provided better agreement with the data for the least defective specimen. The scattering rates were adopted without modification with the
exception of the exponent of the Umklapp scattering rates (we used 10 rather than 12 as above) and the prefactor of $\tau_{4U}^{-1}$
(we decreased it by a factor 380). As noted in Ref.~\onlinecite{ForneyJackle}, these changes put our
four-magnon Umklapp scattering rate in better agreement with that computed by Schwabel and Michel \cite{SchwablMichel}, and produced
better agreement with the data.
With these modifications, the only remaining adjustable parameter was the impurity concentration ($c$).

The scattering rates were incorporated into an interpolation formula for the magnon thermal conductivity using the function described
in Ref.~\onlinecite{deTomas} and derived by Alvarez and Jou \cite{AlvarezJou}:

\begin{equation}
  \kappa_m=\frac{1}{3}C_mv_m\left[\ell_R^B (1-\Sigma)+\ell_R F(L_{eff})\Sigma \right],
\end{equation}

\begin{equation}
\nonumber 
  F(L_{eff})=\frac{1}{2\pi^2}\left(\frac{L_{eff}}{\ell}\right)^2\left(\sqrt{1+4\pi^2\left(\frac{\ell}{L_{eff}}\right)^2}-1\right),
\end{equation}
\noindent where $\Sigma=1/(1+\ell_N/\ell_R)$,
$L_{eff}=\pi\ell_0/(2\sqrt{2})$, $\ell\equiv \sqrt{\ell_N\ell_R}$,
$\ell_R=(1/\ell_{3U}+1/\ell_{4U}+1/\ell_i)^{-1}$, and
$\ell_R^B=(1/\ell_0+1/\ell_R)^{-1}$. We used $\Sigma^2$ in place
of $\Sigma$ in the above expression as it provided a better
interpolation $\to 0$ at low-$T$ (Fig.~\ref{Fig8}).

The impurity scattering concentrations ($c$) employed to produce the curves in Fig.~3 correlate with those found
for phonon-defect scattering (Fig.~\ref{Fig9}) in the Callaway analysis of $\kappa_L$ (Table~I).


\begin{thebibliography}{30}
\bibitem{DipoleNote} Dipole corrections to the spin-wave dispersion are negligible for $k_BT\gg \hbar\omega_m=g\mu_B 4\pi M_s$. For Cu$_2$OSeO$_3$,
$\mu_0 M_s\simeq 0.15$~T yields $\hbar\omega_m\simeq 0.018\ {\rm meV}\simeq 0.21k_B$. Thus the quadratic dispersion should be valid for $T\gg 0.2$~K, and certainly for $T\geq 1$~K.
\end{thebibliography}
\end{document}